\documentclass[aps,prd,twocolumn,showpacs,superscriptaddress]{revtex4}
\usepackage{graphicx}
\usepackage{amsmath}
\usepackage{amssymb,bbm}
\usepackage{multirow}
\usepackage{color}
\usepackage{ulem}
\newcommand{\bq}{\begin{eqnarray}}
\newcommand{\nq}{\end{eqnarray}}

\begin{document}

\title{Predicting Lepton Flavor Mixing from $\Delta(48)$ and Generalized CP Symmetries}

\author{Gui-Jun Ding}
\email{dinggj@ustc.edu.cn}
\affiliation{Department of Modern Physics, University of Science and Technology of China, Hefei, Anhui 230026, China}

\author{Ye-Ling Zhou}
\email{zhouyeling@ihep.ac.cn}
\affiliation{Institute of High Energy Physics, Chinese Academy of
Sciences, P.O. Box 918, Beijing 100049, China}


\begin{abstract}

We propose to understand the mixing angles and CP-violating phases from the $\Delta(48)$ family symmetry combined with the generalized CP symmetry. A model-independent analysis is performed by scanning all the possible symmetry breaking chains. We find a new mixing pattern with only one free parameter, excellent agreement with the observed mixing angles can be achieved and all the CP-violating phases are predicted to take nontrivial values.
This mixing pattern is testable in the near future neutrino oscillation and neutrinoless double-beta decay experiments. Finally, a flavor model is constructed to realize this mixing pattern.
\end{abstract}

\pacs{14.60.Pq, 14.80.Cp, 11.30.Hv}

\maketitle


Discrete family symmetry has been widely used to explain the lepton flavor mixing~\cite{flasy} in the past years. The discovery of a sizable value of $\theta_{13}$ by reactor experiments~\cite{DayaBay} excludes many neutrino mixing models, and opens the possibility of measuring the Dirac CP-violating phase in the next generation neutrino experiments.
The underlying physics of flavor mixing and CP violation is still an open question. The history of physics tells us that symmetry always plays a crucial role in understanding the natural world. Inspired by the success of the family symmetry paradigms, it is natural to extend the family symmetry to include a generalized CP (GCP) symmetry $H_{CP}$~\cite{GCPearly,Feruglio:2012cw, Holthausen:2012dk,Ding:2013hpa}, to predict both flavor mixing angles and CP phases.
This idea has been implemented within $S_4$~\cite{Feruglio:2012cw,Ding:2013hpa,Li:2013jya}, $A_4$~\cite{Ding:2013bpa} and $T'$~\cite{Girardi:2013sza} family symmetries, where the lepton mixing matrix is found to depend on one single parameter $\vartheta$, which can be fixed by the measurement of the mixing angle $\theta_{13}$.

In this paper, we propose to impose the $\Delta(48)$ family symmetry together with GCP symmetry on the theory. Compared with the well-known $S_4$ and $A_4$ family symmetries, $\Delta(48)$ provides many candidates for GCP transformations which could lead to new mixing patterns. After a brief introduction to the GCP symmetry and the group theory of $\Delta(48)$, we derive all the GCP transformations which are consistent with $\Delta(48)$. Then we present possible lepton mixing patterns derived from different symmetry breaking chains in a model-independent way. Finally we focus on phenomenological implications and model building aspect of a new pattern that has not been discussed in the literature. A longer and more complete version of this paper has been presented in \cite{Ding:2014hva}.


A field multiplet $\phi$ transforms under the family and GCP symmetries as
\bq
\phi\stackrel{g}{\longrightarrow} \rho(g)\phi \quad \text{and} \quad
\phi\stackrel{CP}{\longrightarrow}X\phi^{*},
\nq
respectively, where $\rho(g)$ is a representation matrix of the group element $g\in G_f$, and $X\in H_{CP}$ is the GCP transformation matrix. Both of them are unitary matrices. It is nontrivial to combine the family symmetry with the GCP symmetry. The so-called consistence equation has to be satisfied: $X\rho^*(g)X^{-1}=\rho(g')$ for $g,g'\in G_f$~\cite{GCPearly,Feruglio:2012cw, Holthausen:2012dk}.  Furthermore, $X$ maps one group element $g$ into another
element $g'$, consequently $X$ corresponds to an automorphism of $G_f$. It has been established that there is a one-to-one correspondence between the GCP transformation and the automorphism of $G_f$~\cite{Grimus:1995zi}.

In the present work, the family symmetry is chosen to be $G_f=\Delta(48)\cong(Z_4\times Z_4)\rtimes Z_3$, which is a finite subgroup of $SU(3)$ of order $48$ with generators $a$, $c$ and $d$ satisfying
\bq
&\hspace{-0.8cm}&a^3=c^4=d^4=1,\quad cd=dc,\nonumber \\
&\hspace{-0.8cm}&aca^{-1}=c^{-1}d^{-1},\quad ada^{-1}=c,
\label{eq:group_rules}
\nq
where $a$ generates $Z_3$ and $c$, $d$ are generators of $Z_4\times Z_4$.
It belongs to the $\Delta(3n^2)$ series~\cite{Luhn:2007uq} with $n=4$.
Any group element $g\in\Delta(48)$ can be expressed as
$g=a^k c^m d^n$
with $k=0,1,2$ and $m,n=0,1,2,3$.

$\Delta(48)$ has eight irreducible representations:
\begin{itemize}
\item 
Three 1-dimensional (1d) representations $\mathbf{1}$, $\mathbf{1^{\prime}}$, $\mathbf{1^{\prime\prime}}$;
\item 
Five 3d representations $\mathbf{3}$, $\mathbf{\overline{3}}$, $\mathbf{3{}^{\prime}}$, $\mathbf{\overline{3}{}^{\prime}}$ and $\mathbf{\widetilde{3}}$, where $\mathbf{\overline{3}} (\mathbf{\overline{3}{}^{\prime}})$ is the complex conjugate of $\mathbf{3} (\mathbf{3{}^{\prime}})$. The former four are the faithful representations of $\Delta(48)$, while the last one is not.
\end{itemize}
We shall work in the generator $a$ diagonal basis. For the representation $\mathbf{3}$, we choose:
\bq
\small
a=\left(
\begin{array}{ccc}
 1 & 0 & 0 \\
 0 & \omega  & 0 \\
 0 & 0 & \omega ^2
\end{array}
\right)\hspace{-2pt},  ~
c=\frac{1}{3}\left(
\begin{array}{ccc}
 1 & 1-\sqrt{3} & 1+\sqrt{3} \\
 1+\sqrt{3} & 1 & 1-\sqrt{3} \\
 1-\sqrt{3} & 1+\sqrt{3} & 1
\end{array}
\right)\hspace{-2pt},
\label{generators}
\nq
with $\omega=e^{i2\pi/3}$, and the representation matrix of $d$ is given by $d=a^{-1}ca$. Some Kronecker products that will be used later are presented here:
$\mathbf{3}\otimes\mathbf{\overline{3}}=\mathbf{1}\oplus\mathbf{1^{\prime}}\oplus\mathbf{1^{\prime\prime}}\oplus\mathbf{3}{}^{\prime}\oplus\mathbf{\overline{3}}{}^{\prime},~
\mathbf{\overline{3}}\otimes\mathbf{\overline{3}}=\mathbf{3}{}_S\oplus\mathbf{3}{}_A\oplus\mathbf{\widetilde{3}},~
\mathbf{3}{}^{\prime}\otimes\mathbf{3}{}^{\prime}=\mathbf{\overline{3}}{}^{\prime}_S\oplus\mathbf{\overline{3}}{}^{\prime}_A\oplus\mathbf{\widetilde{3}},~
\mathbf{3}\otimes\mathbf{3}{}^{\prime}=\mathbf{3}\oplus\mathbf{3}{}^{\prime}\oplus\mathbf{\widetilde{3}},~
\mathbf{3}\otimes\mathbf{\overline{3}}{}^{\prime}=\mathbf{3}\oplus\mathbf{\overline{3}}{}^{\prime}\oplus\mathbf{\widetilde{3}},~
\mathbf{\overline{3}}{}^{\prime}\otimes\mathbf{\widetilde{3}}=\mathbf{3}\oplus\mathbf{\overline{3}}\oplus\mathbf{3}{}^{\prime}$.

The basic paradigm is that the symmetry $\Delta(48)\rtimes H_{CP}$ is respected at high energy scales, and is then spontaneously broken to different subgroups $G_\nu\rtimes H^{\nu}_{CP}$ and $G_l\rtimes H^l_{CP}$ in the neutrino and charged lepton sectors by flavon fields. This misalignment between the symmetry breaking patterns leads to particular predictions for mixing angles and CP phases. Without loss of generality,
three generations of the left-handed lepton doublets are assigned to $\Delta(48)$ triplet $\mathbf{3}$.
The invariance of the Lagrangian under residual family symmetries and residual GCP symmetries implies that the neutrino mass matrix
$m_{\nu}$ and the charged lepton mass matrix $m_{l}$ should satisfy
\begin{subequations}
\begin{eqnarray}
&\label{eq:residual_f}\hspace{-1cm}& \rho^{\dag}(g_{\nu})m_{\nu}\rho^*(g_{\nu})=m_{\nu}, ~~
\rho^{\dagger}(g_{l})m_{l}m^{\dagger}_{l}\rho(g_{l})=m_{l}m^{\dagger}_{l},\\
&\label{eq:residual}\hspace{-1cm}& X^{\dag}_{\nu}m_{\nu}X^*_{\nu}=m^{*}_{\nu}, \quad \hspace{0.8cm}
X^{\dagger}_{l}m_{l}m^{\dagger}_{l}X_{l}=(m_{l}m^{\dagger}_{l})^{*},
\label{eq:residual}
\end{eqnarray}
\end{subequations}
where neutrinos are assumed to be Majorana particles, $g_{\nu}$, $g_{l}$ denote the group elements of the residual family symmetries $G_{\nu}$, $G_l$, and $X_{\nu}$, $X_{l}$ denote the elements of the remnant GCP symmetries $H^{\nu}_{CP}$, $H^{l}_{CP}$, respectively.

By systematically scanning all the possible remnant family subgroups $G_{\nu}$ and $G_{l}$, we find that only the case $G_{\nu}=Z_2$ and $G_{l}=Z_3$ can lead to viable phenomenology. One can choose $G_{\nu}=\{1,c^2\}$ and $G_{l}=\{1,a, a^2\}$ without loss of generality, since all the possible choices are related by group conjugation.
From the constraint of Eq.~\eqref{eq:residual_f}, we find that the charged lepton mass matrix $m_l$ is diagonal in the chosen basis, and the neutrino mass matrix $m_{\nu}$ takes the form:
\bq
\hspace{-0.8cm}
m_{\nu}&=&\alpha\left(
\begin{array}{ccc}
 2 & -1 & -1 \\
 -1 & 2 & -1 \\
 -1 & -1 & 2
\end{array}
\right)+\beta\left(
\begin{array}{ccc}
 1 & 0 & 0 \\
 0 & 0 & 1 \\
 0 & 1 & 0
\end{array}
\right)\nonumber\\
\hspace{-0.8cm}&+& \gamma\left(
\begin{array}{ccc}
 0 & 1 & 1 \\
 1 & 1 & 0 \\
 1 & 0 & 1
\end{array}
\right)+\epsilon\left(
\begin{array}{ccc}
 0 & 1 & -1 \\
 1 & -1 & 0 \\
 -1 & 0 & 1
\end{array}
\right),
\label{eq:massmatrx}
\nq
where $\alpha$, $\beta$, $\gamma$ and $\epsilon$ are complex parameters, and they are further constrained by the neutrino residual GCP symmetry $H^{\nu}_{CP}$, as shown in Eq.~\eqref{eq:residual}.

\begin{table*}[tb]
\begin{center}
\begin{tabular}{ p{3.5cm} p{3cm} p{3cm} p{3.5cm} p{4cm} }
\hline\hline

  &  \text{\tt pattern A}  &    \text{\tt pattern B}   & \text{\tt pattern C} &  \text{\tt pattern D} \\  \hline

\text{\tt GCP matrix} $X$ & $\rho(c^{2k_1+k_2}d^{2k_2})P_{23}$ & $\rho(c^{2k_1+k_2}d^{2k_2})$ & $\rho(c^{2k_1+k_2+1}d^{2k_2})P_{23}$ & $\rho(c^{m}d^{2k_2+1})P_{23}$ \\\hline

$\sin^2\theta_{13}$ & $\frac{1}{3}-\frac{1}{3}\cos\vartheta$ & $\frac{1}{3}-\frac{1}{3}\cos\vartheta$ & $\frac{1}{3}-\frac{1}{2\sqrt{3}}\cos\vartheta$ & $\frac{1}{3}-\frac{1+\sqrt{3}}{6\sqrt{2}}\cos\vartheta$ \\

$\sin^2\theta_{12}$ & $\frac{1}{2+\cos\vartheta}$ & $\frac{1}{2+\cos\vartheta}$ & $\frac{2}{4+\sqrt{3}\cos\vartheta}$ & $\frac{2 \sqrt{2}}{4\sqrt{2}+\left(1+\sqrt{3}\right) \cos\vartheta}$ \\

$\sin^2\theta_{23}$ & $\frac{1}{2}$ & $\frac{1}{2}\mp\frac{\sqrt{3}\sin\vartheta}{4+2\cos\vartheta}$ & $\frac{1}{2}\mp\frac{\sqrt{3}\cos\vartheta}{8+2\sqrt{3}\cos\vartheta}$ & $\frac{1}{2}\mp\frac{\left(3-\sqrt{3}\right)\cos\vartheta}{8\sqrt{2}+2\left(1+\sqrt{3}\right) \cos\vartheta}$ \\

$J_{CP}$ & $-\frac{\sin\vartheta}{6\sqrt{3}}$ & 0 & $\mp\frac{\sin\vartheta}{6 \sqrt{3}}$ & $\mp\frac{\sin\vartheta}{6\sqrt{3}}$ \\

$\left|\tan\delta\right|$ & $+\infty$ & 0 & $\big|\frac{4+\sqrt{3}\cos\vartheta}{1+\sqrt{3}\cos\vartheta}\tan\vartheta\big|$ & $\big|\frac{4\sqrt{2}+\left(1+\sqrt{3}\right) \cos\vartheta}{1-\sqrt{3}-\sqrt{2}\cos\vartheta}\tan\vartheta\big|$ \\

$\left|\tan\alpha_{21}\right|\,\text{or}\,\left|\cot\alpha_{21}\right|$  & 0 & 0 & $\big|\frac{\sqrt{3}+2\cos\vartheta}{\sin\vartheta}\big|$ & $\big|\frac{1+\sqrt{3}+2\sqrt{2}\cos\vartheta+\left(1-\sqrt{3}\right) \sin\vartheta}{1+\sqrt{3}+2\sqrt{2}\cos\vartheta-\left(1-\sqrt{3}\right) \sin\vartheta }\big|$ \\

$\left|\tan\alpha^{\prime}_{31}\right|$ & 0 & 0 & $\big|\frac{4\sqrt{3}\sin\vartheta}{1-3\cos2\vartheta}\big|$ & $\big|\frac{4 \sin\vartheta}{2-3\sqrt{3}+\left(2+\sqrt{3}\right)\cos2\vartheta}\big|$ \\
\hline\hline
\end{tabular}
\caption{\label{tab:mixingparameters} The predictions for lepton mixing patterns and the associated mixing parameters for all possible choices of residual GCP symmetries in the neutrino sector, and all the mixing patterns are found to depend on only one parameter $\vartheta$ varying from $0$ to $2\pi$. The related GCP matrices hold for all faithful 3d representations $\mathbf{3}$, $\mathbf{\overline{3}}$, $\mathbf{3{}^{\prime}}$ and $\mathbf{\overline{3}{}^{\prime}}$ with  $k_1,k_2=0,1$, $m=0,1,2,3$. The sign $``+\infty"$ for $\left|\tan\delta\right|$ implies that the corresponding Dirac CP-violating phase is $\pm\pi/2$. }
\end{center}
\end{table*}

Each GCP transformation corresponds to an automorphism of the family symmetry $G_f$. 
The automorphism group of $\Delta(48)$ is $\mathrm{Aut}(\Delta(48))\cong\Delta(48)\rtimes D_8$ with 384 group elements.
Its outer automorphism group is proven to be a dihedral group $\mathrm{Out}(\Delta(48))\cong D_8$, with generators $u_1$ and $u_2$ defined as
\bq
\left\{\begin{array}{l}
a\stackrel{u_1}{\longrightarrow}a^2\\
c\stackrel{u_1}{\longrightarrow}cd^2
\end{array}\right.,\qquad
\left\{\begin{array}{l}
a\stackrel{u_2}{\longrightarrow}a\\
c\stackrel{u_2}{\longrightarrow}cd^2
\end{array}\right..
\nq
The following multiplication rules are fulfilled
\bq
u^4_1=u^2_2=\left(u_1u_2\right)^2=id\,.
\nq
Each group element in $\mathrm{Out}(\Delta(48))$ can be expressed as $u_1^\mu u_2^\nu$ for $\mu=0, 1, 2, 3$ and $\nu=0, 1$. The generators $u_1$ and $u_2$ act on the irreducible representation of $\Delta(48)$ as
\begin{eqnarray}
\nonumber&\hspace{-1cm}&\mathbf{1^{\prime}}\stackrel{u_1}{\longleftrightarrow}\mathbf{1^{\prime\prime}},~ \mathbf{3}\stackrel{u_1}{\longrightarrow}\mathbf{3^{\prime}}\stackrel{u_1}{\longrightarrow}\mathbf{\overline{3}}\stackrel{u_1}{\longrightarrow}\mathbf{\overline{3}^{\prime}}\stackrel{u_1}{\longrightarrow}\mathbf{3}, ~ \mathbf{\widetilde{3}}\stackrel{u_1}{\longrightarrow}\mathbf{\widetilde{3}}\,,\\
&\hspace{-1cm}&\mathbf{1^{\prime}}\stackrel{u_2}{\longrightarrow}\mathbf{1^{\prime}},~ \mathbf{1^{\prime\prime}}\stackrel{u_2}{\longrightarrow}\mathbf{1^{\prime\prime}},~ \mathbf{3}\stackrel{u_2}{\longleftrightarrow}\mathbf{3^{\prime}},~ \mathbf{\overline{3}}\stackrel{u_2}{\longleftrightarrow}\mathbf{\overline{3}^{\prime}},~ \mathbf{\widetilde{3}}\stackrel{u_2}{\longrightarrow}\mathbf{\widetilde{3}}\,.
\end{eqnarray}

The 8 outer automorphisms generated by $u_1$ and $u_2$ lead to different CP transformations and should have distinct physical implications.  In the present work, we minimally extend the $\Delta(48)$ family symmetry to include only those nontrivial CP transformations which map one irreducible representation into its complex conjugate. We find that there are three outer automorphisms, $u_1^2$, $u_1u_2$, and $u_1^3u_2$, satisfying this requirement.
\begin{itemize}
\item
The first automorphism $u_1^2$ interchanges all 3d irreducible representations with their complex conjugate representations. The corresponding GCP matrix in each 3d irreducible representation is determined to be just a permutation $X(u^2_1)=P_{23}$. Thus, the GCP transformation acts on a 3d field $\phi=(\phi_1,\phi_2,\phi_3)^T$ as
\bq
\left(\begin{array}{ccc}
\phi_1 \\
\phi_2 \\
\phi_3
\end{array}\right)
\stackrel{CP}{\longrightarrow}
P_{23}\left(\begin{array}{ccc}
\phi_1^* \\
\phi_2^* \\
\phi_3^*
\end{array}\right)=
\left(\begin{array}{ccc}
\phi_1^* \\
\phi_3^* \\
\phi_2^*
\end{array}\right).
\nq
This is the so-called $\mu-\tau$ reflection symmetry~\cite{mutau}.
\item
The second automorphism $u_1u_2$ interchanges $\mathbf{3}{}^{\prime}$ with $\mathbf{\overline{3}}{}^{\prime}$ but maps $\mathbf{3}$ and $\mathbf{\overline{3}}$ into themselves. 
In $\mathbf{3}{}^{\prime}$ and $\mathbf{\overline{3}}{}^{\prime}$ representation space, we find it to be $X(u_1u_2)=\mathbbm{1}_3$, i.e., the conventional CP transformation
\bq
\left(\begin{array}{ccc}
\phi_1 \\
\phi_2 \\
\phi_3
\end{array}\right)
\stackrel{CP}{\longrightarrow}
\mathbbm{1}_3\left(\begin{array}{ccc}
\phi_1^* \\
\phi_2^* \\
\phi_3^*
\end{array}\right)=
\left(\begin{array}{ccc}
\phi_1^* \\
\phi_2^* \\
\phi_3^*
\end{array}\right).
\nq
\item
The third one $u_1^3u_2$ exchanges $\mathbf{3}$ with $\mathbf{\overline{3}}$ while maps $\mathbf{3}{}^{\prime}$ and $\mathbf{\overline{3}}{}^{\prime}$ into themselves. In $\mathbf{3}$ and $\mathbf{\overline{3}}$, the corresponding GCP matrix is $X(u^3_1u_2)=\mathbbm{1}_3$ as well.
\end{itemize}
Taking account of the inner automorphisms, we find that residual GCP transformations compatible with the remnant family symmetry $G_{\nu}=\{1, c^2\}$ can be expressed as
\bq
X_\nu=\rho(c^md^n)P_{23} ~\text{or} ~\rho(c^md^n)
\label{eq:GCPmatrix}
\nq
in all 3d representations with $m,n=0,1,2,3$.


We shall further investigate the constraints on the neutrino mass matrix $m_\nu$ in Eq.~\eqref{eq:massmatrx} by applying the residual GCP symmetry and then derive the PMNS matrix.
In the PDG convention~\cite{pdg}, the PMNS matrix is cast in the form
\bq
U_{PMNS}=V\,\text{diag}(1,e^{i\frac{\alpha_{21}}{2}},e^{i\frac{\alpha_{31}}{2}})\,,
\label{eq:pmns_pdg}
\nq
with
\begin{equation}\small
V\hspace{-1.5pt}=\hspace{-1.5pt}\left(\begin{array}{ccc}
\hspace{-3pt}c_{12}c_{13}  \hspace{-2pt}&\hspace{-2pt}   s_{12}c_{13}  \hspace{-2pt}&\hspace{-2pt}   s_{13}e^{-i\delta}\hspace{-3pt} \\
\hspace{-3pt}-s_{12}c_{23}-c_{12}s_{23}s_{13}e^{i\delta}  \hspace{-2pt}&\hspace{-2pt}  c_{12}c_{23}-s_{12}s_{23}s_{13}e^{i\delta} \hspace{-2pt}&\hspace{-2pt}  s_{23}c_{13}\hspace{-3pt}  \\
\hspace{-3pt}s_{12}s_{23}-c_{12}c_{23}s_{13}e^{i\delta}  \hspace{-2pt}&\hspace{-2pt} -c_{12}s_{23}-s_{12}c_{23}s_{13}e^{i\delta} \hspace{-2pt}&\hspace{-2pt}  c_{23}c_{13}\hspace{-3pt}
\end{array}\right)\hspace{-3pt},
\end{equation}
in which $c_{ij}=\cos\theta_{ij}$, $s_{ij}=\sin\theta_{ij}$, $\delta$ is the Dirac CP-violating phase and $\alpha_{21}$, $\alpha_{31}$ are the Majorana CP-violating phases. It is more convenient to redefine the Majorana phase $\alpha^{\prime}_{31}\equiv\alpha_{31}-2\delta$ during the analysis of the neutrinoless double-beta decay.

With different choices of remnant GCP transformations in Eq.~\eqref{eq:GCPmatrix} and abandoning the cases which predict degenerate neutrino masses, we obtain 4 kinds of mixing patterns, denoted by patterns A, B, C and D. Each mixing pattern depends on one free parameter $\vartheta$ and predicts $\sin^2\theta_{12}=1/(3\cos^2\theta_{13})$ since the structure of $m_\nu$ in Eq.~\eqref{eq:massmatrx} preserves the second column of the PMNS matrix as $(1/\sqrt{3},1/\sqrt{3},1/\sqrt{3})^T$. As a consequence, mixing angles as well as CP phases are strongly correlated, as shown in Table~\ref{tab:mixingparameters}. For proper values of $\vartheta$, all cases are compatible with the present neutrino oscillation data~\cite{GonzalezGarcia:2012sz} within $3\sigma$ range, except $\theta_{13}$ in pattern C.

\begin{figure}[h!]
\begin{center}
\includegraphics[width=0.48\textwidth]{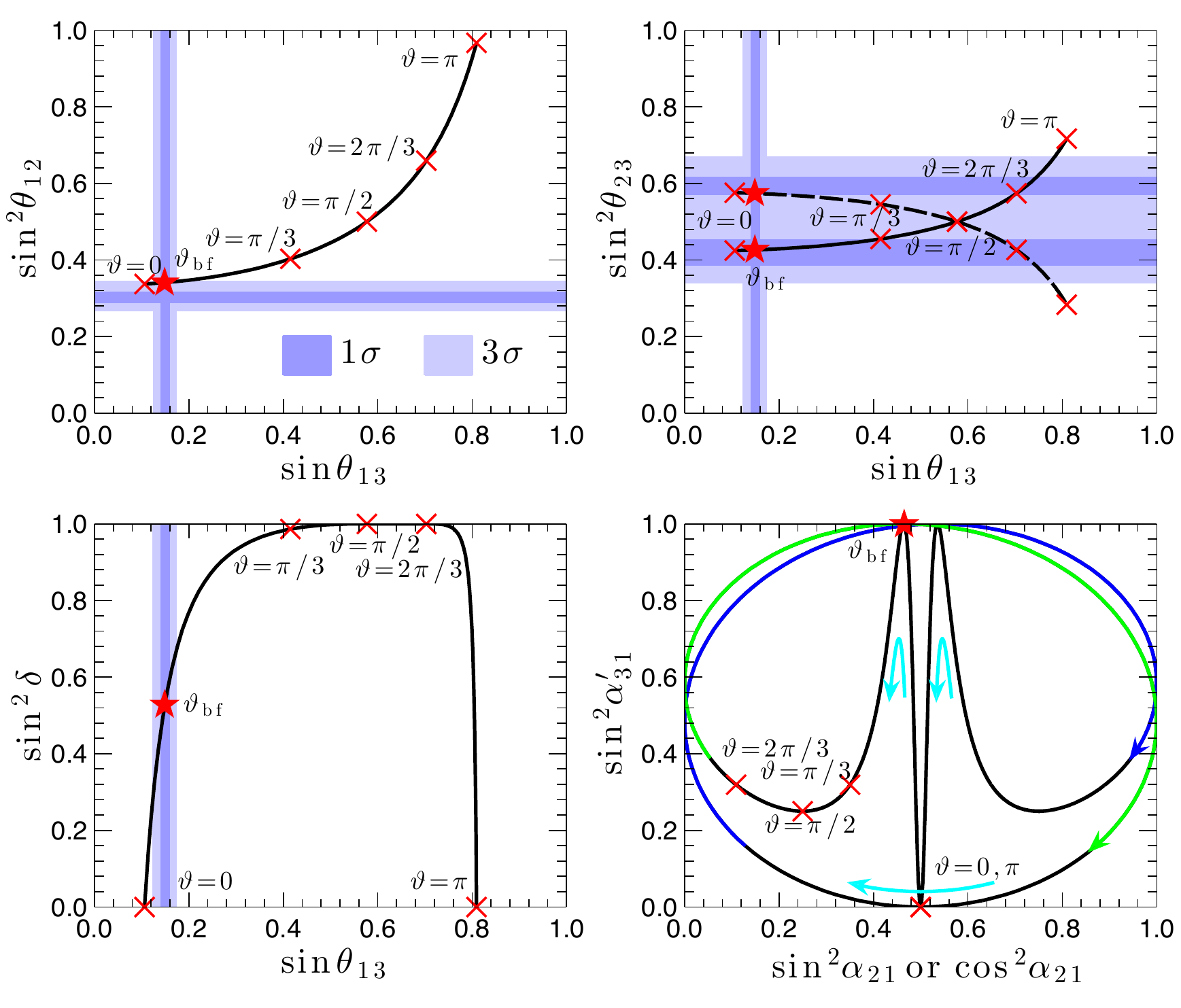}
\caption{\label{fig:patternD} Correlations among the mixing angles and CP-violating phases in pattern D. We mark the best-fit value $\theta_{\rm bf}$ of the parameter $\vartheta$ with a red star, and also label $\vartheta=0,\pi/3,\pi/2,2\pi/3,\pi$ with a red cross on the curve. In the top right panel, the results of $\sin^2\theta_{23}$ for the first octant and the second octant of $\theta_{23}$, are shown in a solid line and dashed line, respectively. The $1\sigma$ and $3\sigma$ ranges of the mixing angles are taken from Ref.~\cite{GonzalezGarcia:2012sz}.}
\end{center}
\end{figure}
\begin{figure}[h!]
\begin{center}
\includegraphics[width=0.45\textwidth]{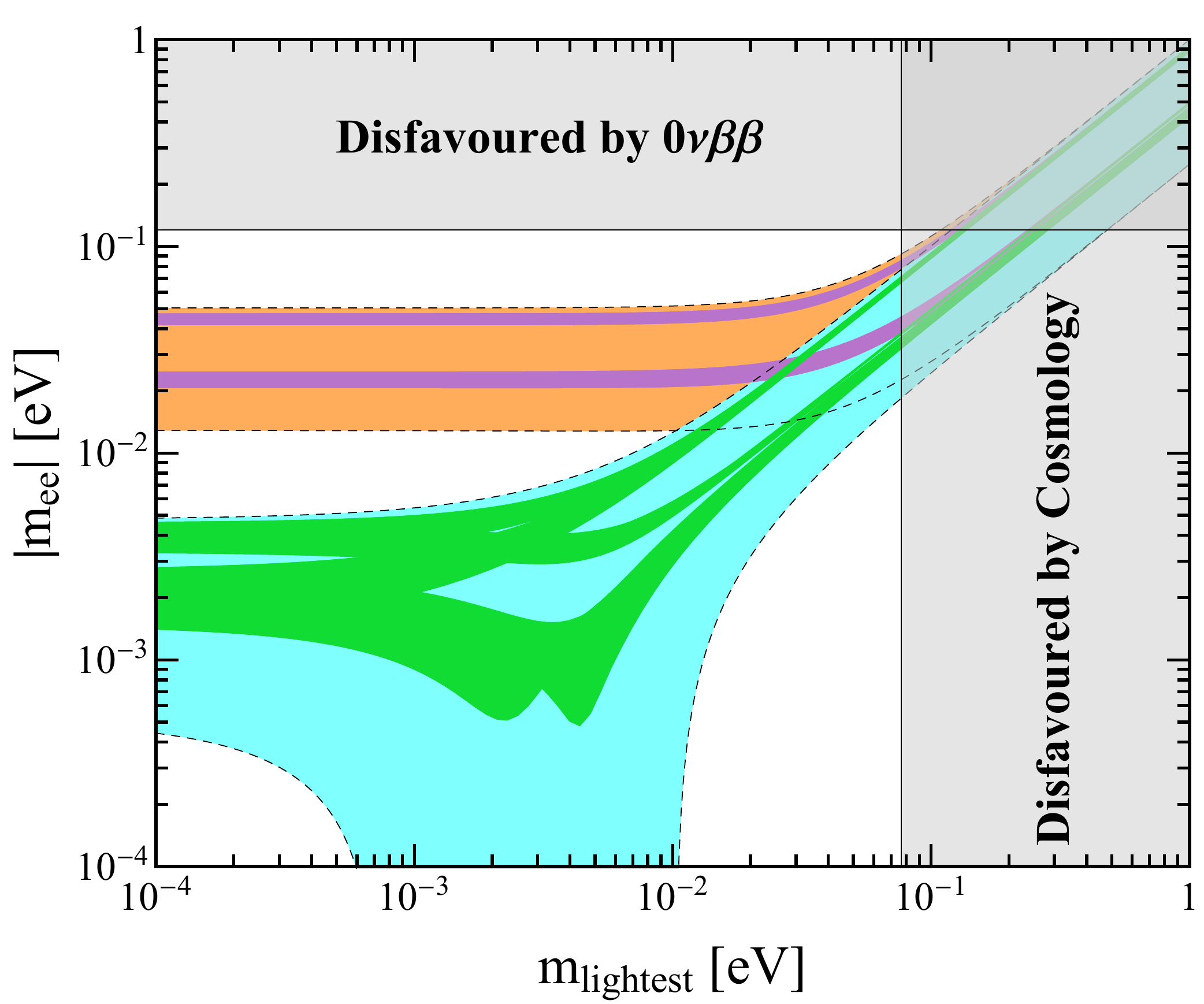}
\caption{\label{fig:meepatternD} The prediction of the effective neutrino mass $\langle m\rangle_{ee}$ as a function of the lightest neutrino mass $m_1$ in normal ordering (NO) or $m_3$ in inverted ordering (IO) in pattern D. The splits within each mass ordering come from the uncertainties of the quadrants of CP-violating phases. }
\end{center}
\vspace{-0.3cm}
\end{figure}

In the following, we will focus on pattern D which is completely new as far as we know.  In this case, the GCP symmetry corresponding to the outer automorphism $u^2_1$ should be implemented. All the mixing parameters, in particular the CP phases are nontrivially dependent on $\vartheta$, and the correlations between the mixing parameters are plotted in Fig.~\ref{fig:patternD}.
Excellent agreement with the present global-fitting data of mixing angles can be achieved. It is interesting to note that the relation between $\sin^2\alpha_{21}$ (or $\cos^2\alpha_{21}$) and $\sin^2\alpha'_{31}$, shown in low right panel, looks like the ``compound eyes'' of an insect.
Taking the $3\sigma$ ranges of mixing angles from~\cite{GonzalezGarcia:2012sz}, we obtain
\bq
0.162\leqslant|\vartheta|\leqslant 0.341,
\label{eq:varthetadata}
\nq
and the CP-violating phases are constrained to lie in the following intervals:
\bq
&\hspace{-0.8cm}&0.292\leqslant\sin^2\delta\leqslant0.667,\quad 0.781\leqslant\sin^2\alpha^{\prime}_{31}\leqslant1,\nonumber\\
&\hspace{-0.8cm}&0.455\leqslant\sin^2\alpha_{21} ~\text{or}~ \cos^2\alpha_{21}\leqslant0.478.
\nq
The quadrants of CP-violating phases cannot be determined in the present model-independent approach. Notice that the Dirac phase $\delta$ is large but not maximal, and this prediction could be tested in the next generation neutrino oscillation experiments LBNE and Hyper-K.

This pattern is also testable in the future neutrinoless double-beta ($0\nu\beta\beta$) decay experiments. The rate of $0\nu\beta\beta$ decay is determined by the nuclear matrix element and the effective parameter $\langle m\rangle_{ee}=|m_1c_{12}^2c_{13}^2+m_2s_{12}^2c_{13}^2e^{i\alpha_{21}}+m_3s_{13}^2e^{i\alpha'_{31}}|$.
In Fig.~\ref{fig:meepatternD}, we show the prediction for $\langle m\rangle_{ee}$ as a function of the lightest neutrino mass, where the constraint in Eq.~\eqref{eq:varthetadata} has been taken into account. The upper bounds from cosmology (the sum of neutrino masses $\sum m_i<0.23$ eV)~\cite{Ade:2013zuv} and the current $0\nu\beta\beta$ bound ($\langle m\rangle_{ee}<0.32$ eV)~\cite{0nu2beta} are also included in the figure.
The next generation $0\nu\beta\beta$ decay experiments will reach the sensitivity of $\langle m\rangle_{ee}\simeq(0.01-0.05)$ eV after 5 years of data taking~\cite{0nu2beta}. As a consequence, if the signal of $0\nu\beta\beta$ would not be observed, the inverted mass ordering scenario of this pattern would be excluded, since we have $\langle m\rangle_{ee}>0.02$ eV in this case as shown in Fig.~\ref{fig:meepatternD}.


Finally, we shall construct a simple flavor model in which pattern D is realized. The field arrangement is listed in Table~\ref{tab:field}, where $\ell_L,e_R,\mu_R,\tau_R$ denote the left-handed and right-handed lepton fields, $H$ represents the Higgs field, and $\phi_l$, $\varphi_l$, $\rho_l$, $\varphi$, $\xi$ are the gauge-singlet flavon fields. The additional $Z_3\times Z_4\times Z_5$ symmetry is used to eliminate undesired dangerous operators and derive suitable vacuum alignments. Yukawa couplings invariant under $\Delta(48)\times Z_3 \times Z_4\times Z_5$ are
\bq
-\mathcal{L}_\ell &\hspace{-4pt}=\hspace{-4pt}&
\frac{y_\tau}{\Lambda} \phi_l \overline{\ell_L}H \tau_R
+\frac{y_\mu}{\Lambda^2} (\phi_l\varphi_l)_\mathbf{3} \overline{\ell_L}H \mu_R \nonumber\\
&\hspace{-4pt}+\hspace{-4pt}&
\frac{y_{e_1}}{\Lambda^3} \big(\rho_l(\phi_l\phi_l)_{\mathbf{\widetilde{3}}}\big)_{\mathbf{3}} \overline{\ell_L}H e_R 
\hspace{-1pt}+\hspace{-1pt}
\frac{y_{e_2}}{\Lambda^3} \big(\phi_l(\varphi_l\varphi_l)_{\overline{\mathbf{3}}{}^{\prime}_S}\big)_{\mathbf{3}} \overline{\ell_L}H e_R \nonumber\\
&\hspace{-4pt}+\hspace{-4pt}&\frac{y_\varphi}{\Lambda^2} \varphi (\overline{\ell_L}\widetilde{H}\widetilde{H}^T\ell^c_L)_{\mathbf{3}_S} 
\hspace{-3pt}+\hspace{-1pt}
\frac{y_\xi}{\Lambda^2} \xi(\overline{\ell_L}\widetilde{H}\widetilde{H}^T\ell^c_L)_\mathbf{\widetilde{3}} +h.c.,
\label{eq:Yukawa_coupling}
\nq
in which $\widetilde{H}=i\sigma_2 H^*$,
and $\Lambda$ is the cut-off scale.
Moreover, all coupling coefficients are real since the GCP symmetry is imposed. 
The flavon vacuum expectation values can be realized by using the supersymmetric driving field method. Here we directly list them as
\bq
&\hspace{-0.5cm}&\langle\phi_l\rangle=v_{\phi_l}(1,0,0)^T,
\langle\varphi_l\rangle=v_{\varphi_l}(0,1,0)^T,
\langle\rho_l\rangle=v_{\rho_l}(0,1,0)^T, \nonumber\\
&\hspace{-0.5cm}&\langle\varphi\rangle=e^{i\frac{\pi}{4}}v_\varphi(1,1, 1)^T, \quad
\langle\xi\rangle=v_\xi(0,-\omega^2,\omega)^T,
\nq
where $v_{\phi_l}$, $v_{\varphi_l}$ and $v_{\rho_l}$ are generally complex, while $v_\varphi$ and $v_\xi$ are real. Notice that the vacuum of the neutrino flavons $\varphi$ and $\xi$ preserve $Z_2\rtimes H^\nu_{CP}$ symmetry, where the residual GCP matrix is $X_\nu=\rho(d)P_{23}$. As a result, pattern D is naturally produced.

\begin{table}[t!]
\begin{center}
\begin{tabular}{ c c c c c c c c c c c c c}
\hline\hline

~{\tt Fields}~  &  ~$\ell_L$~  & ~$e_R$~ & ~$\mu_R$~ & ~$\tau_R$~ & \,\,$H$\,\, &  ~$\phi_l$~ & ~$\varphi_l$~  &  ~$\rho_l$~ &  \,~$\varphi$~\, & ~$\xi$~  \\  \hline

$\Delta(48)$ & $\mathbf{3}$ & $\mathbf{1}$ & $\mathbf{1}$ & $\mathbf{1}$ & $\mathbf{1}$ & $\mathbf{3}$ & $\mathbf{3}'$ & $\mathbf{\overline{3}}{}^{\prime}$ & $\mathbf{\overline{3}}$ & $\mathbf{\widetilde{3}}$ \\

$Z_3$ & $\omega^2$ & $\omega$ & $1$ & $\omega^2$ & $1$ & $1$ & $\omega^2$ & $\omega$ & $\omega$ & $\omega$ \\

$Z_4$ & $-i$ & 1 & $i$ & $-1$ & $1$ & $i$ & $i$ & $i$ & $-1$ & $-1$ \\

$Z_5$ & $1$ & $\omega_5^2$ & $1$ & $\omega_5^3$ & $1$ & $\omega_5^2$ & $\omega_5^3$ & $\omega_5^4$ & $1$ & $1$ \\

\hline\hline
\end{tabular}
\caption{\label{tab:field}Fields and their transformation properties under $\Delta(48)\times Z_3 \times Z_4\times Z_5$, where $\omega_5=e^{i2\pi/5}$ and $\ell_L=(\ell_{\tau L},\ell_{\mu L},\ell_{e L})^T$.}
\end{center}
\end{table}

Leptons acquire masses after symmetry breaking. The charged lepton mass matrix is found to be diagonal with
\bq
&\hspace{-1cm}&m_e=\Big|y_{e_1} \frac{v_{\rho_l}v_{\phi_l}^2}{\Lambda^3}v
+2\omega y_{e_2} \frac{v_{\phi_l}v_{\varphi_l}^2}{\Lambda^3}v\Big|, \nonumber\\
&\hspace{-1cm}&m_\mu=\Big|y_\mu \frac{v_{\phi_l}v_{\varphi_l}}{\Lambda^2} v\Big|, ~~
m_\tau=\Big|y_\tau \frac{v_{\varphi_l}}{\Lambda} v\Big|,
\nq
in which $v=\langle H\rangle=175$ GeV.
The neutrino mass matrix is of the form of Eq.~\eqref{eq:massmatrx} with
\bq
\hspace{-0.3cm}\alpha=e^{i\frac{\pi}{4}}\frac{y_\varphi v_\varphi v^2}{\Lambda^2},
\beta=-\omega^2\frac{y_\xi v_\xi v^2}{\Lambda^2},
\gamma=-\epsilon=\omega \frac{y_\xi v_\xi v^2}{2\Lambda^2}.
\nq
The PMNS matrix is exactly pattern D, and the parameter $\vartheta$ fulfills $\tan \vartheta=y_\xi v_\xi/(2\sqrt{3}\,y_\varphi v_\varphi)$. For the neutrino masses, we find that the neutrino mass spectrum can only be NO. A detailed calculation shows that
\bq
\frac{\Delta m^2_{21}}{\Delta m^2_{31}}=\frac{1}{2}-\frac{3\sin^2\vartheta-1}{4\sin\vartheta}\,.
\nq
To be compatible with data $|\Delta m^2_{21}/\Delta m^2_{31}|\simeq 0.03$, we find $\vartheta\simeq-0.351$, or equivalently, $y_\xi v_\xi\simeq -1.27y_\varphi v_\varphi$, which leads to the predictions:
\bq
&\hspace{-0.8cm}&\theta_{13}\simeq 10.14^\circ, ~\theta_{12}\simeq35.9^\circ,~\theta_{23}\simeq40.8^\circ,\nonumber\\
&\hspace{-0.8cm}&\delta\simeq304.4^\circ, ~\alpha_{21}\simeq222.3^\circ, ~\alpha_{31}\simeq352.9^\circ\,.
\nq
The lightest neutrino mass $m_1$ and the effective mass $\langle m\rangle_{ee}$ is fixed in this model,
\bq
m_1\simeq 0.0278 ~\text{eV}, \quad
\langle m\rangle_{ee}\simeq0.0112~ \text{eV}.
\nq

In summary, we have proposed the discrete group $\Delta(48)$ to explain lepton mixing angles and predict CP-violating phases in the framework of generalized CP symmetries. $\Delta(48)$ has a large automorphism group and thus provides rich choices for GCP transformations. By systematically scanning all the possible symmetry breaking chains, we find 4 different mixing patterns compatible with experimental data. Among them, pattern D is a completely new mixing pattern that has not been discussed in the literature. It predicts nontrivial CP-violating phases, which can be tested in the future neutrino oscillation and neutrinoless double-beta decay experiments. We have realized this pattern in an effective flavor model, where all the neutrino flavor mixing parameters, the absolute scale of neutrino masses and $\langle m\rangle_{ee}$ are fixed.

We would like to thank Z.Z. Xing for simulating discussion, continuous support, and reading the manuscript. We are also grateful to Y.F. Li for his helpful discussion. This work was
supported in part by the National Natural Science Foundation of
China under Grant Nos. 11275188, 11179007, and 11135009.

\end{document}